\documentclass[preprint,12pt]{elsarticle}
\usepackage{amssymb}
\usepackage{graphicx}

\journal{Nuclear Physics B}

\begin{document}

\begin{frontmatter}

\title{An idea for detecting {\it capture} dominated Dark Stars}
\author{Fabio Iocco}
\address{Institut de Physique Th\'eorique, CNRS, URA 2306 \& CEA/Saclay, F-91191 Gif-sur-Yvette, France\\
Institut d`Astrophysique de Paris, UMR 7095-CNRS Paris, Universite' Pierre et Marie Curie, boulevard Arago 98bis, 75014, Paris, France\\
email address: iocco@iap.fr}

\begin{abstract}
I discuss an idea which could lead
to a methodology for testing the
effects of WIMP DM scattering and capture
onto primordial stars. 
It relies on the effects of ``life prolongation''
of affected PopIII stars, that can slow down
nuclear reactions by supporting their own 
structure with the 
energy produced by annihilating DM
captured inside the star. 
This can lead to an alteration of
the Pair Production SuperNova rate, which
could constitute a peculiar signature of the
existence of {\it capture} Dark Stars.
\end{abstract}

\begin{keyword}
Early universe, dark matter, population III stars 
\end{keyword}
\end{frontmatter}

\begin{center}
{\bf Prolegomena}
\end{center}
In the last two years there has been considerable activity 
to understand the effects of Weakly Interacting Massive Particles 
(WIMP) dark matter (DM) on the first stars to form
in a $\Lambda$CDM Universe, the Population III (PopIII). 
PopIII affected by WIMPs are commonly referred to as ``Dark Stars'';
in the following I will use this name for  sake of simplicity,
although preceded by {\it gravitational} and {\it capture} (dominated) 
to avoid any misunderstanding between completely different phenomena,
that rely on diverse properties of the DM particle.
In this Section, I will briefly summarize the state of the art of the field
and identify the key-issues relevant to detection possibilities. 
Readers are addressed to the original references for a thorough 
discussion of the parameter dependence and  details about the 
values adopted.

PopIII are thought to form by the very smooth collapse of the first  
${\cal O}$(10$^6$M$_\odot$) DM and gas haloes to
undergo non--linear collapse at redshifts z$\lesssim$30.
Several groups, working at simulations with different
techniques agree in describing the PopIII stellar formation as
very different from local, low-redshift one, with remarkable effects such
as the formation of a single, massive star per halo at its very center.
Such objects are thought to live short lives  -- $\tau_*\sim$10$^6$yr --
and, depending on their mass, either directly collapse 
to a Black Hole or explode as Pair Production SuperNovae (PPSN), 
up to 100 times more powerful than normal typeII SN;
for a review of the state of the art of the field, see e.g \cite{FS3proc}.
Weakly Interacting Massive Particles (WIMPs) are well
motivated dark matter particle candidates, being the
Lightest Supersymmetric Partner of the standard model, 
see e.g. \cite{DMrevs} for reviews.
Stable in models with R--parity conservation,
they bear the remarkable properties to be
self--annihilating and weakly coupled to standard model
particles.

In \cite{SFG}, the authors first showed that following the
gas cooling and collapse, DM can concentrate in the center 
forming a steeper than NFW profile (to which I will refer in the
following using the term ``cusp''), and that this conjures in such a way that 
(for a wide range of DM masses and self-annihilation cross--sections)
the heating induced by DM annihilation can overcome the feeble cooling
of the pre--stellar cloud. 
The authors speculated this could generate a long--lasting
equilibrium phase, and named the resulting astrophysical object a {\it Dark Star};
in the following I will refer to the condition in which gas cooling and DM heating 
rates equal as  ``transition point''.
This phase involves no scattering of DM off the gas: 
the WIMPs are gravitationally contracted by the gas collapse, and self--annihilate
thus constituting an (additional) energy source; these
effects are more efficient at late times during star formation, but
yet far from the actual proto--star assembling. The results
of \cite{Natarajan:2008db}, based on gas and DM profiles from
three--dimensional simulations (without DM feedback on the chemistry or
the equation of state of the system) confirmed the results of \cite{SFG} up to the reach of the 
transition point, falling below simulation resolution afterwards.

In \cite{RipaIDM}, the authors have studied the collapse phase of the gas 
preceding the transition point, investigating whether non--linear effects induced 
by the heating following DM annihilation could modify the properties of the gas
before getting to it. They concluded that although DM annihilation
{\it does} influence the properties of the gas even at intermediate states, 
changes are not so dramatic to modify the Jeans mass.
Although this intermediate phase clearly deserves more study, 
and the behavior of the gas cooling and collapse in presence
of DM annihilation need to be better understood, current results
hint that the transition point can be reached
without dramatic changes in presence
of feedback between DM annihilation and the gas chemistry.

Once the formation of a proto--stellar object is achieved,
the problem becomes to understand 
the evolution of an hydrostatic gas structure
sustained by an underlying profile of self--annihilating DM, 
gravitationally coupled to the baryons; this has been addressed
in \cite{Iocco:2008rb, Freese:2008wh, Spolyar:2009nt}
with different approaches. As  of now, the results of the two 
groups seem to be discrepant: the authors of 
\cite{Freese:2008wh, Spolyar:2009nt} find that the DM
concentrated inside the hydrostatic structure can sustain it
against gravitational collapse for times of ${\cal O}$ (10$^6$yr),
thus letting the seed object accrete several hundreds of solar masses
before igniting nuclear reactions; according to this scenario the main
outcome of the gravitational Dark Star mechanism is a population
of supermassive objects of ${\cal O}$ (800M$_\odot$).
The authors of  \cite{Iocco:2008rb} do instead find a much less pronounced
characteristic feature: a gravitational Dark Star is delayed in its
normal path toward the Zero Age Main Sequence of up to no more
than 10$^5$yr. This finding is obtained in absence of any
gas accretion onto the seed object, it is however clear that even including
accretion at the same rates used by \cite{Freese:2008wh, Spolyar:2009nt},
the mass of the final object cannot be altered dramatically; this is 
also confirmed by recent findings of the authors of  \cite{Iocco:2008rb},
so far released in several seminars and which will soon made public.
Here I will rely on the latter scenario, in which the
final outcome of the gravitational Dark Star mechanism,
is a population not much dissembling the standard one, and 
slightly more massive, depending on parameters. 
\begin{center}
{\bf Capture phase}
\end{center}
Before studying the effects of the gravitational
phase on the baryonic structure, it was also realized by \cite{CaptDisc}
that the peculiar formation of a PopIII star at the bottom of the
gravitational potential -at the center of the formed DM cusp- 
would dramatically enhance the effects of 
WIMP capture (and subsequent annihilation) on the star, 
would it eventually form. 
It is to be noticed that this is a very different mechanism from 
the {\it gravitational} contraction, and it relies on weak interactions
between baryons and WIMPs {\it external} to the star; 
a thorough review of the physics involved in this mechanism
can be found in \cite{Scott:2008ns},
whereas a more qualitative description of the differences between
the two processes is summarized in \cite{Iocco:2008ps}. Capture effects
are irrelevant in early stages of evolution (and during the gravitational phase),
typically starting to become relevant when the star is already on the Hayashi track,
see \cite{Iocco:2008rb}.
The effects of WIMP capture from an external reservoir 
on the evolution of the first  metal-free stars were studied
in \cite{Iocco:2008rb,Yoon:2008km,Taoso:2008kw} and 
the results of the different groups are in remarkable agreement.
The outcome can be summarized as follows:
{\it i)} the lifetime of the star is prolonged (virtually up to higher than
a Hubble time)  in presence of high enough DM densities 
{\it outside} the star (environmental), 
as DM ``fueling'' slows down nuclear reactions 
and supports the star;  
{\it ii)} the time--prolonging effect is a very sharp function of
the environmental DM density times the
elastic scattering WIMP--baryon cross section, see e.g. 
Fig. 3 in \cite{Taoso:2008kw}, thus permitting to define a 
critical DM density (once picked the baryon-DM scattering cross--section value);
{\it iii)} the final fate of the star (SN/BH) is unchanged, mainly because
of the little effects of capture during the late stages of stellar
evolution, see \cite{Yoon:2008km}.
\begin{center}
{\bf An idea for detecting {\it capture} dominated Dark Stars}
\end{center}
The idea I propose relies on all of the properties mentioned 
above: a single PopIII star fueled by {\it captured} DM will live
as long as the external DM density will stay above the 
critical DM density (see 
Fig. 6 in \cite{Iocco:2008rb} for a dependence of the critical DM density
as a function of stellar mass), or the star itself moves out of the DM reservoir. 
In \cite{Yoon:2008km} we estimated this time, 
$\Delta\tau_\chi$, as of ${\cal O}$ (10$^8$yr), the typical 
timescale of a major halo merger at redshifts $z\sim$10, relevant for
PopIII stars. I will adopt this number 
--taken as a value averaged over the population-- 
for practical purposes and comment more about it later.
As a consequence of the life--prolongement, capture
dominated Dark Stars will contribute differently than a
standard PopIII to Reionization and stellar formation
feedback in the early Universe; the consequences of this
in terms of detectability of Dark Stars has been
addressed with a first dedicated study by \cite{Schleicher:2008gk}.
Another remarkable fact is that $\Delta\tau_\chi$
is much bigger than the typical lifetime of massive stars, 
$\tau_*\sim$10$^6$yr, and of the Hubble time at z$\sim$10.
Therefore, whereas in a standard scenario the birth and death
of massive PopIII are instantaneous on cosmological 
timescales, they will differ of $\Delta\tau_\chi$ in a capture
Dark Star one.
\begin{figure}[t]
\centering
 \includegraphics[angle=0,width=0.8\textwidth]{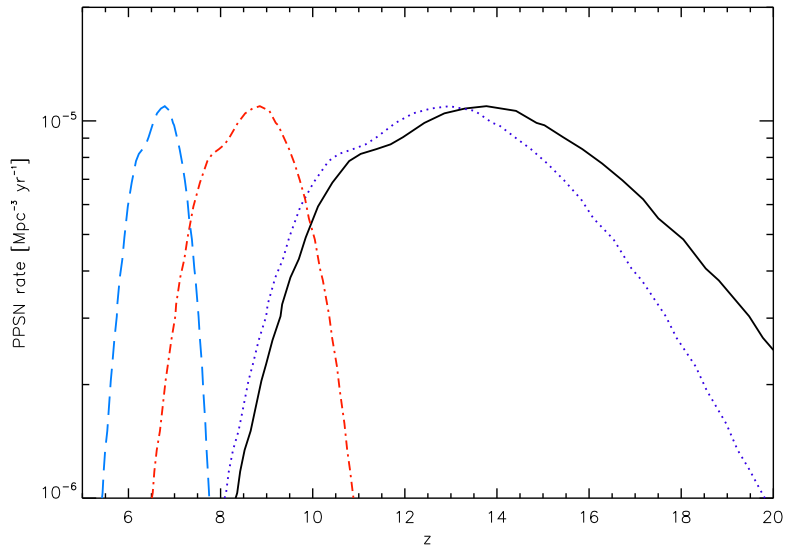}
\caption{Shift of the PPSN rate as a consequence of a 
``life prolongement'' effect due to DM capture on massive PopIII stars.
The standard rate (solid line) is obtained by Fig. 3 in Choudhury \& Ferrara (2005),
the dotted, dot-dashed, and dashed lines are for
$\Delta\tau_\chi$=5$\times$10$^7$, 5$\times$10$^8$,10$^9$yr,
respectively.}
\label{PPSNshift}
\end{figure}
In terms of population, this may result in a delay of the
PPSN rate, which will be more appreciable at higher redshift, 
where the ratio between $\Delta\tau_\chi$ and the Hubble time is bigger.
In Figure \ref{PPSNshift}, I show the effects of different $\Delta\tau_\chi$ on
a PPSN rate. For this simple exercise I have taken the PopIII stellar
formation rate (SFR) from the self--consistent reionization
models from \cite{CFreio} and converted it into a PPSN rate
by assuming a monochromatic initial mass function (IMF),
M$_*$=150M$_\odot$; conversion of different SFRs and 
IMFs is trivial, and the shift effect does not change qualitatively.
The most remarkable features are the absence (or drastic reduction) of PPSN event 
until redshifts corresponding to Hubble times bigger than $\Delta\tau_\chi$,
and the appearance of events at lower redshifts, where there wouldn't be 
any in the standard case. 
However, the detectability of these features will be troubling:
whereas the first relies on the capability to observe Pair Production
SN at extremely high redhisfts, 
the latter, of intrinsic easier accessibility, would require (even more than the first) 
a detailed knowledge of the PPSN rate at low redshifts.  
Several authors have estimated it, 
see e.g. \cite{PPSNrate}, but the results obtained
are extremely sensitive to the properties of the 
Population III (as of now yet unaccessible with direct observations)
and the quoted numbers vary of orders of magnitude.
Hopefully we will have gathered more and enough knowledge 
to develop a methodology based on this idea by the JWST era,
and then it will be time to discuss its feasibility and the
disentanglement of the information; however, it is 
possible to briefly comment on some intrinsic constraints to its basis
with elements at our hand as of now.
In particular, $\Delta\tau_\chi$ will depend and be limited by:
{\it i)} the self annihilation time of the DM cusp on which the star 
is ``sitting''; {\it ii)} the depletion of an overcritical DM density by 
the star's own capture and use; {\it iii)} the deplacement of the star 
with respect to the cusp or the cusp disruption for friction, tidal effects
etc; the latter can take place as a consequence of a merger, and its
typical timescale at z$\sim$10 is of ${\cal O}$ (10$^8$yr).
Concerning the self-annihilation timescale of the DM cusp, by
taking a $\rho_{DM}\sim$10$^{11}$GeV/cm$^3$ (of the order of
the critical densities for massive stars, by assuming a spin-dependent 
elastic scattering cross-section with baryons 
$\sigma_0$=5$\times$10$^{-39}$cm$^2$), and non exotic WIMP
parameters such as mass m$_\chi$=100GeV/c$^2$ and
$\langle\sigma v \rangle$=3$\times$10$^{-26}$cm$^3$/s, one gets 
$\tau_{sa}\sim$10$^{10}$yr; this is  the timescale on which the self--annihilation 
of the DM cusp {\it outside} the star will bring it below the critical density
for life--prolonging effects on the star.
The star also contributes to DM depletion by using it instead of nuclear 
burning to sustain itself: in 
one-dimensional models based on the Adiabatic Contraction, 
one typically finds that approximately ${\cal O}$ (1M$_\odot$) of 
DM is concentrated at supercritical densities around the star; for
objects of M$_*$$\sim$150M$_\odot$ with characteristic luminosities
of ${\cal O}$ (10$^6$L$_\odot\sim$10$^{39}$erg/s) this reservoir
takes approximately $\tau_*^\chi\sim$10$^8$yr to be exhausted or lowered
to undercritical densities (in principle accretion can bring more DM in the
region surrounding the star, however the clue-issue is that the
DM density has to stay supercritical to have the dramatic 
life-prolongement effect required for this framework).
This back-of-the-envelope calculation based on
central parameter values, show that $\Delta\tau_\chi$
has an upper limit of approximately 10$^{10}$yr, whereas a more
reasonable value seems to lie in the range 10$^8$yr; our knowledge of
this parameter will definitely benefit from  dedicated 
high resolution simulations of the central regions of a primordial
star forming halo, 
as well as of the behavior of a central baryonic object and of the DM
during mergers of small halos at high redshifts.

I acknowledge fruitful conversations with
A.~Ferrara and M.~Kamionkowski.


\begin{thebibliography}{00}

\bibitem{FS3proc} 
  V.~Bromm, N.~Yoshida, L.~Hernquist and C.~F.~McKee,
  Nature {\bf 459}, 49 (2009);
"First Stars III", 2008, AIP Conf.~Proc., 990, T.~Abel, A.~Heger, and B.~W.~`O.~Shea eds.
\bibitem{DMrevs} 
  G.~Jungman, M.~Kamionkowski and K.~Griest,
  Phys.\ Rept.\  {\bf 267}, 195 (1996);
L. Bergstrom, Rep.\ Prog.\ Phys.\ 63, 793  (2000);
  G.~Bertone, D.~Hooper and J.~Silk,
  Phys.\ Rept.\  405, 279 (2005).
\bibitem{SFG}
  D.~Spolyar, K.~Freese and P.~Gondolo,
  Phys.\ Rev.\ Lett.\ 100, 051101 (2008).
\bibitem{Natarajan:2008db}
  A.~Natarajan, J.~C.~Tan and B.~W.~O'Shea,
  Astrophys.\ J.\  {\bf 692}, 574 (2009).
\bibitem{RipaIDM}
  E.~Ripamonti, F.~Iocco, A.~Bressan, R.~Schneider, A.~Ferrara and P.~Marigo,
  PoS(idm2008)075,  arXiv:0903.0346 [astro-ph.CO].
\bibitem{Iocco:2008rb}
  F.~Iocco, A.~Bressan, E.~Ripamonti, R.~Schneider, A.~Ferrara and P.~Marigo,
  Mon.\ Not.\ Roy.\ Astron.\ Soc.\  {\bf 390}, 1655 (2008).
\bibitem{Freese:2008wh}
  K.~Freese, P.~Bodenheimer, D.~Spolyar and P.~Gondolo,
  Astrophys.\ J.\  {\bf 685}, L101 (2008).
\bibitem{Spolyar:2009nt}
  D.~Spolyar, P.~Bodenheimer, K.~Freese and P.~Gondolo,
  arXiv:0903.3070 [astro-ph.CO].
\bibitem{CaptDisc} 
  F.~Iocco,
  Astrophys.\ J.\  {\bf 677}, L1 (2008);
  K.~Freese, D.~Spolyar and A.~Aguirre,
  JCAP {\bf 0811}, 014 (2008).
\bibitem{Scott:2008ns}
  P.~Scott, M.~Fairbairn and J.~Edsjo,
  Mon.\ Not.\ Roy.\ Astron.\ Soc.\  {\bf 394}, 82 (2009).
\bibitem{Iocco:2008ps}
  F.~Iocco, A.~Bressan, E.~Ripamonti, R.~Schneider, A.~Ferrara and P.~Marigo,
 Procs. IAU Symposium 255 (2008)  arXiv:0809.2417 [astro-ph].
\bibitem{Yoon:2008km}
  S.~C.~Yoon, F.~Iocco and S.~Akiyama,
  Astrophys.\ J.\  {\bf 688}, L1 (2008).
\bibitem{Taoso:2008kw}
  M.~Taoso, G.~Bertone, G.~Meynet and S.~Ekstrom,
  Phys.\ Rev.\  D {\bf 78}, 123510 (2008).
\bibitem{Schleicher:2008gk}
  D.~R.~G.~Schleicher, R.~Banerjee and R.~S.~Klessen,
  Phys.\ Rev.\  D {\bf 79}, 043510 (2009).
\bibitem{CFreio}
  T.~R.~Choudhury and A.~Ferrara,
  Mon.\ Not.\ Roy.\ Astron.\ Soc.\  {\bf 361}, 577 (2005);
  T.~R.~Choudhury and A.~Ferrara,
  Mon.\ Not.\ Roy.\ Astron.\ Soc.\  {\bf 371}, L55 (2006).
 \bibitem{PPSNrate} 
   S.~M.~Weinmann and S.~J.~Lilly,
  Astrophys.\ J.\  {\bf 624}, 526 (2005);
  J.~H.~Wise and T.~Abel,
  Astrophys.\ J.\  {\bf 629}, 615 (2005);
  E.~Scannapieco, P.~Madau, S.~Woosley, A.~Heger and A.~Ferrara,
  Astrophys.\ J.\  {\bf 633}, 1031 (2005).
        
\end{thebibliography}
\end{document}